\documentclass[twocolumn,showpacs,preprintnumbers,amsmath,amssymb]{revtex4}

\usepackage{graphicx}
\usepackage{dcolumn}
\usepackage{bm}
\usepackage{graphicx}
\usepackage{epsfig}
\include{graphic}

\begin{document}
\preprint{APS/123-QED}
\title{Realization of a Room-Temperature Spin Dynamo: The Spin Rectification Effect}

\author{Y. S. Gui, N. Mecking, X. Zhou, G. Williams, and C. -M. Hu\footnote{Electronic address: hu@physics.umanitoba.ca}}

\affiliation{Department of Physics and Astronomy, University of
Manitoba, Winnipeg, Canada R3T 2N2}

\date{\today}

\begin{abstract}

We demonstrate a room temperature spin dynamo where the precession
of electron spins in ferromagnets driven by microwaves manifests
itself in a collective way by generating d.c. currents. The
current/power ratio is at least three orders of magnitude larger
than that found previously for spin-driven currents in
semiconductors. The observed bipolar nature and intriguing
symmetry are fully explained by the spin rectification effect via
which the nonlinear combination of spin and charge dynamics
creates d.c. currents.

\end{abstract}


%
%
%
%
%
%

\maketitle There is currently great interest in generating d.c
currents via spin dynamics \cite{Tulapurkar,Sankey,Ganichev,Yang}.
The significance is two fold: on the one hand, it provides
electrical means for investigating spin dynamics, while on the
other hand, it may pave the way for designing new spin sources for
spintronic applications. In semiconductor materials with
spin-orbit coupling, both aspects have been demonstrated
\cite{Ganichev,Yang,Hu}. Progress with ferromagnetic metals (FM)
has been achieved using magnetic multilayers
\cite{Tulapurkar,Sankey}. Very recently, a few groups
\cite{Gui,Costache} have begun to develop techniques for
electrical detection of spin resonances in a FM single layer.
Understanding the d.c. effects in a FM single layer is crucial for
clarifying whether spin pumping effects might exist in magnetic
multilayers \cite{Costache,Azevedo}. Until now, in contrast to the
case for semiconductors \cite{Ganichev,Yang}, the important
question of how to effectively generate d.c currents from a single
FM remains unclear. This leaves our understanding of the interplay
between spin dynamics and electrostatic response incomplete, as
evidenced in the controversial views represented in recent work
\cite{Costache,Azevedo}.

In this paper, we present both experimental and theoretical
answers to this question. A spin dynamo is constructed which
generates d.c currents via spin waves in a FM microstrip. The
observed bipolar nature and intriguing symmetry allow us to
unambiguously identify the origin of this phenomenon as the spin
rectification effect. A general formula is obtained which fully
describes our experimental results. To give a simple picture of
the spin rectification effect, we begin with the well-known
optical rectification effect, which occurs in nonlinear media with
large second order suspetibility. Here the optical response to the
product of time-dependent electric fields $e_{0}cos(\omega t)$ is
governed by the trigornometric relation: $cos(\omega_{1}t) \cdot
cos(\omega_{2}t)=\{cos[(\omega_{1}-\omega_{2})t]+cos[(\omega_{1}+\omega_{2})t]\}/2$.
If the frequencies $\omega_{1}=\omega_{2}$, the terms with
difference and sum frequencies causes optical rectification and
second harmonic generation, respectively. Similar nonlinear
dynamic response to the product of RF electric and magnetic fields
is the origin of the spin rectification effect, investigated here
by using a spin dynamo.

\begin{figure} [t]
\begin{center}
\epsfig{file=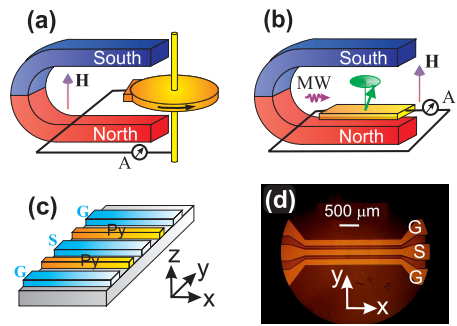,height=4.2 cm} \caption{(color online). (a)
Faraday's dynamo with a revolving copper disk converts energy from
rotation to a current of electricity. (b) Spin dynamo with a FM
strip converts energy from spin precession to a bipolar current of
electricity. (c) Diagram of the spin dynamo structure with Py
strips placed in slots between the ground (G) and signal (S) lines
of a coplanar waveguide. (d) Top view micrography of a
device.}\label{Fig.1}
\end{center}
\end{figure}

The spin dynamo we constructed is sketched in Fig. 1b and c. It is
the microscopic counterpart of Faraday's dynamo shown in Fig. 1a.
Both devices generate d.c currents in a static magnetic field, but
the rotational motion of a macroscopic copper plate in the
Faraday's dynamo is replaced by microscopic spin precession in the
spin dynamo. The device is based on a permalloy (Py) microstrip
(typically 2.45 mm $\times$ 20 $\mu$m $\times$ 137 nm) placed in
the slot between the ground (G) and signal (S) strips of a
coplanar waveguide (CPW) \cite{Wen}. From anisotropic magneto
resistance (AMR) and ferromagnetic resonance (FMR) measurement,
the following material parameters have been determined for the Py
strip: the conductivity $\sigma$ = 3 $\times$ 10$^{4}$
$\Omega^{-1}cm^{-1}$, the magneto anisotropy $\Delta \rho/\rho$ =
0.019, the saturation magnetization $\mu_{0}M_{0}$ = 1.0 T, and
the demagnetization factors $N_{x} \approx$ 0, $N_{y} \approx$
0.007, $N_{z} \approx$ 0.993. Here, $\mu_{0}$ is the permeability
of the vacuum. The coordinate system is shown in Fig. 1c. The CPW
is deposited with Au/Ag/Cr layers (5/550/5 nm) on top of a semi
insulating GaAs substrate and is impedance matched to 50 $\Omega$.
The dimensions of the CPW are 150 $\mu$m and 100 $\mu$m in width
for the strips and the slots, respectively. As sketched in Fig.
1b, setting the device in an electromagnet and feeding the CPW
with microwaves using a conventional microwave power generator,
d.c. currents are generated in the Py strip at room temperature.
To preserve the symmetry of the CPW, two identical Py microstrips
are inserted in both slots. They are measured independently, and
the same effect is found. Several spin dynamos with different Py
thickness have been measured in various configurations. The data
reported here have been rendered, both experimentally and
theoretically, to convey the most significant aspects of the
observed phenomena. For the same purpose, a special sample holder
has been constructed that enables the spin dynamo to rotate about
both $x$ and $y$-axes with an angular resolution approaching 0.01
degree.

Fig. 2 demonstrates the production of bipolar d.c. currents from
the Py strip under microwave radiation. The microwave frequency is
set at 5.4 GHz. The current $I$ flowing along the $x$-axis is
measured using a current amplifier while sweeping the magnetic
field $H$ applied nearly perpendicular to the Py strip. We define
$\alpha$ and $\beta$ as the small angles of the field direction
tilted away from the $z$-axis towards $x$ and $y$ axes,
respectively. As shown in Fig. 2a, when $\beta$ is set to zero,
the current $I(\alpha,H)$ measured at $\alpha$ = -1$^{\circ}$
shows a positive peak and a negative dip. The current rapidly
diminishes when $\alpha$ is tuned to zero, and then changes
polarity when $\alpha$ becomes positive (Fig. 2b). The same
bipolar behaviour holds true for the current $I(\beta,H)$ measured
at $\alpha$ = 0 and plotted in Figs. 2c and d, except that its
polarity also changes upon reversing the direction of the applied
magnetic field. In addition, the sharp features in Figs. 2a and b
disappear in Figs. 2c and d. The insets in Fig. 2 summarize the
angular dependence of the maximum bipolar current, measured by
using standard lock-in techniques to enhance the signal/noise
ratio at extremely small angles. The curious results of Fig.2 can
be summarized with the following observed bipolar symmetry:
\begin{eqnarray}
I(\alpha, H) = -I(-\alpha, H) = I(\alpha, -H),~for~\beta =0; \cr
I(\beta, H) = -I(-\beta, H) = -I(\beta, -H),~for~\alpha =0.
\end{eqnarray}

\begin{figure} [t]
\begin{center}
\epsfig{file=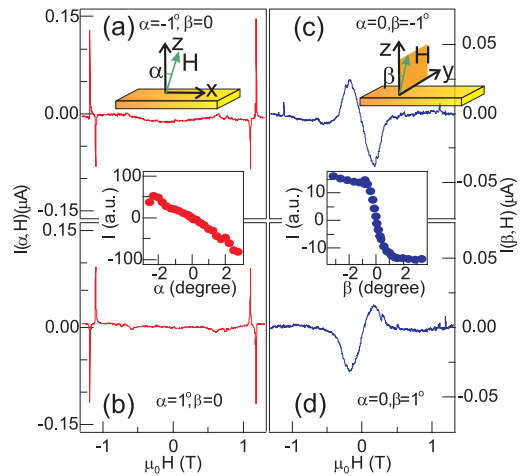,height=6.2 cm} \caption{(color online).
Bipolar d.c. currents measured as a function of the magnetic field
$H$ applied with angles (a) $\alpha$ = -1$^{\circ}$, $\beta$ =0,
(b) $\alpha$ = 1$^{\circ}$, $\beta$ =0, (c) $\alpha$ = 0, $\beta$
= -1$^{\circ}$, and (d) $\alpha$ = 0, $\beta$ = 1$^{\circ}$. The
microwave frequency is fixed at 5.4 GHz. Insets summarize the
angular dependence of the maximum current measured at $H <$ 0.}
\label{Fig.2}
\end{center}
\end{figure}

The sharp features in the $I(\alpha,H)$ trace observed in Figs. 2a
and b suggest a resonant nature, which is confirmed by frequency
dependence measurements. At higher frequencies, up to four
resonances are observed. For $|H|>M_{0}$, the measured resonant
relations, plotted in Fig. 3a, follow the dispersion of standing
spin waves (SSW) in Py films \cite{Morrish}, given by
$\omega=\gamma (|H|-M_{0}+2Ak^{2}/\mu_{0}M_{0})$. Here, $k=p
\pi/d$ is the wave vector with the values of $p$ determined by the
number of half wavelength in the Py strip with a thickness $d$.
The solid lines in Fig. 3a are calculated using a gyromagnetic
ratio $\gamma=181\mu_{0}$ GHz/T and an exchange constant $A$ =
1.22 $\times$ 10$^{-11}$ N for Py \cite{Morrish}. Four modes with
$p$ = 0, 2, 3, and 4 are determined from the observed resonances,
which indicates an intermediate pinning condition
\cite{Puszkarski}. With $|H|<M_{0}$, the dispersion for SSW was
previously unclear, due partially to the experimental challenge of
detecting spin waves in a single Py microstrip. However, a similar
field curve with a sharp cusp corresponding to the magnetic
anisotropy was observed for the FMR \cite{Vonsovskii}. Here, in
the situation with $\beta$ = 0, $\mathbf{M}$ rotates with
increasing $H$ from the easy axis parallel to the $x$-axis towards
the direction of $\mathbf{H}$. When $\alpha$ is small, we find the
solution of the magnetostatic problem gives $cos\varphi \approx H
cos\alpha/M_{0}$ where the angle $\varphi$ shown in Fig. 3a
describes the direction of $\mathbf{M}$ with respect to the hard
$z$-axis. This simple relation is confirmed by the AMR measured at
$\beta$ = 0, plotted in Fig. 3b, which is well represented by
$R(H)=R(\infty)[1+(\Delta\rho/\rho)sin^{2}\varphi]$. By noticing
the apparent similarity between the measured SSW dispersion and
the AMR trace shown in Figs. 3a and b, respectively, we suggest an
empirical expression describing SSW at $|H|<M_{0}$, given by
$\omega=2\gamma(Ak^{2}/\mu_{0}M_{0})\sqrt{1+(4\pi/p)sin^{2}\varphi}$,
which well describes the measured dispersions.

It follows therefore that the spin dynamo is an ideal device for
electrically detecting spin resonances in FM microstructures. In
order to focus on the origin of the spin rectification effect, we
leave the interesting physics of spin waves in the spin dynamo to
a later paper and continue here by studying $I(\beta,H)$ measured
at $\alpha$ = 0. Since $\mathbf{H}$ in this case is tilted towards
the $y$-axis by a small angle $\beta$, and because $N_{x} \ll
N_{y} \ll N_{z}$ for the long Py strip, we make an approximation
to simplify the otherwise complicated magnetostatic problem. With
increasing $H$, we assume $\mathbf{M}$ rotates first from the easy
$x$-axis towards the $y$-axis, before it moves out of the $xy$
plane towards the direction of $\mathbf{H}$. Physically this means
the Py strip is much easier to magnetize along the $y$-axis than
the $z$-axis. By describing the direction of $\mathbf{M}$ with the
angle $\theta$ measured with respect to the $y$-axis as shown in
Fig. 3c, we find $cos\theta =H/M_{1}$ for $|H|<M_{1}$, and $\theta
\approx$ 0 for $|H|>M_{1}$, where $M_{1}\equiv
N_{y}M_{0}/sin\beta$. As shown in Fig. 3b, our approximation is
justified by the AMR trace measured at $\alpha$ = 0 and $\beta
\approx$ -1$^{\circ}$, which agrees well with the curve calculated
from $R(H)=R(\infty)[1+(\Delta\rho/\rho)sin^{2}\theta]$ for
$|H|<M_{1}\approx$ 0.4 T. In this field range, we find that the
current $I(\beta,H)$ partially follows $\partial R(H)/\partial H$.
This comparison is shown in Fig. 3c.

All of these results indicate that the current is induced by the
dynamics of $\mathbf{M}$, the equilibrium direction of which is
determined by the angles $\alpha$ and $\beta$ at $|H|>M_{0}$, and
$\varphi$ and $\theta$ at $|H|<M_{0}$. These results imply that
the electric and magneto response in the spin dynamo are coupled,
as demonstrated by the following arguments on the nature of the
spin rectification effect.

\begin{figure} [t]
\begin{center}
\epsfig{file=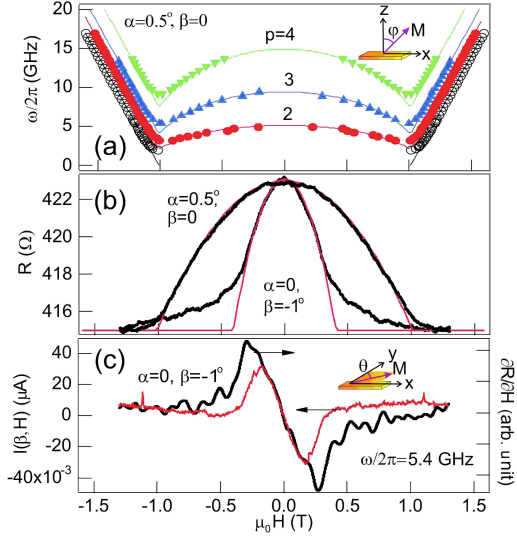,height=7.2 cm} \caption{(color online). (a)
Frequency variation of the resonant features in the $I(\alpha,H)$
trace measured at $\alpha$ = 0.5$^{\circ}$, $\beta$ = 0 (marks).
The data can be well described by the dispersions (solid curves)
for SSW. (b) AMR trace $R(H)$ measured (thick curves) and
calculated at $\alpha$ = 0.5$^{\circ}$, $\beta$ = 0 and $\alpha$ =
0, $\beta$ = -1$^{\circ}$. (c) Current trace $I(\beta,H)$ measured
at $\alpha$ = 0, $\beta$ = -1$^{\circ}$ and 5.4 GHz (thin curve)
in comparison with the field derivative of the AMR trace $\partial
R(H)/\partial H$ (thick curve).} \label{Fig.3}
\end{center}
\end{figure}

It is well known that under internal magnetic ($\mathbf{H}_{i}$)
and electric ($\mathbf{E}$) fields, the magnetostatic response of
the Py strip is determined by
$\mathbf{M}=\hat{\chi_{0}}\mathbf{H}_{i}$ via the static
permeability tensor $\hat{\chi_{0}}$, while the electrostatic
response is described \cite{Jan} by the generalized Ohm's equation
$\mathbf{J}=\sigma\mathbf{E}-R_{A}(\mathbf{J}\cdot
\mathbf{M})\mathbf{M}+\sigma R_{0}\mathbf{J}\times
\mathbf{H}_{i}+\sigma R_{1}\mathbf{J}\times \mathbf{M}$. This
equation takes into account spin-charge coupling effects
phenomenogically via the nonlinear terms. Here the second term
describes magnetoanisotropy via the AMR coefficient
$R_{A}=\Delta\rho/\rho M_{0}^{2}$, which we have used to calculate
the AMR trace $R(H)$. The last two terms describe the Hall effect
with the ordinary and extraordinary Hall coefficient given by
$R_{0}$= -1.9 $\times$ 10$^{-8}$ $\Omega$cm/T and $R_{1}\approx$
3.3 $\times$ 10$^{-8}$ $\Omega$cm/T, respectively \cite{Foner} .
Dynamically, the magneto and electro responses of the Py strip to
the RF magnetic ($\mathbf{h}$) and electric ($\mathbf{e}$) fields
are given by $\mathbf{m}=\hat{\chi}\mathbf{h}$ and
$\mathbf{j}=\hat{\sigma}\mathbf{e}$, respectively. Here, the
high-frequency permeability and conductivity tensors $\hat{\chi}$
and $\hat{\sigma}$ determine the dynamic spin accumulation
$\mathbf{m}$ and the eddy current $\mathbf{j}$, respectively. If
static and dynamic fields coexist, the nonlinear effects couple
not only the magneto and electric response, but also the static
and dynamic ones. Solving the equations for both static and
dynamic response self consistently, we obtain the d.c. current
density $\mathbf{I}$ in the absence of the applied static electric
field as
\begin{equation}
\mathbf{I}=\mathbf{J}_{0}-R_{A}(\mathbf{J}_{0}\cdot\mathbf{M})\mathbf{M}+\sigma
R_{0}\mathbf{J}_{0}\times\mathbf{H}_{i}+\sigma
R_{1}\mathbf{J}_{0}\times\mathbf{M}
\end{equation}
where $\mathbf{J}_{0}=-R_{A}[\langle\mathbf{j}\times
\mathbf{m}\rangle\times
\mathbf{M}+\langle\mathbf{j}\cdot\mathbf{m}\rangle\mathbf{M}]+\sigma
R_{0}\langle\mathbf{j}\times\mathbf{h}\rangle+\sigma
R_{1}\langle\mathbf{j}\times\mathbf{m}\rangle$ is determined by
the time average \cite{Juretschke} of the product of $\mathbf{h}$
and $\mathbf{e}$.

Eq. (2) is the general expression for the spin-rectification
effect in FM. Taking into account the large electric and magneto
anisotropy of the Py microstrip, the measured current density in
our spin dynamo is obtained:
\begin{equation}
I\approx-2R_{A}M_{x}\{\langle j_{x}m_{x}\rangle-\sigma
R_{1}[M_{y}\langle j_{x}m_{z}\rangle-M_{z}\langle
j_{x}m_{y}\rangle]\}
\end{equation}

Since the components of $\mathbf{M}$ are described by the angles
$\alpha$, $\beta$, $\varphi$ and $\theta$ defined before, it is
straight forward to show that if $\beta$ = 0,
$I(\alpha,H)\approx-2R_{A}M_{0}\langle j_{x}m_{x}\rangle
sin\alpha$ for $|H|>M_{0}$, and
$I(\alpha,H)\approx-2R_{A}M_{0}\sqrt{1-(H/M_{0})^{2}}\langle
j_{x}m_{x}\rangle(\alpha/|\alpha|)$ for $|H|<M_{0}$. When $\alpha$
= 0, $I(\beta,H)=0$ for $|H|>M_{0}$, and
$I(\beta,H)\approx2(R_{A}/N_{y})\langle j_{x}m_{y}\rangle
Hsin\beta$ for $|H|<M_{1}$. From these results, the bipolar
symmetry deduced for $I(\alpha,H)$ and $I(\beta,H)$ is exactly the
same as summarized in Eq. (1). The result $I(\beta,H)=0$ for
$|H|>M_{0}$ explains the disappearance of the sharp resonances in
Figs. 2c and d. Note that for $|H|<M_{1}$, $I(\beta,H)$ is
quasi-resonant which combines the resonant nature of $m_{y}$ and a
nonresonant linear dependence on $H$. It is easy to show that
$|I(\beta,H)|$ saturates at about $2R_{A}\langle j_{x}m_{y}\rangle
M_{0}$ at small angles and $I(\beta,H)$ follows $\partial
R(H)/\partial H$ in the field dependence, which agree well with
the experimental results shown in Figs. 2 and 3, respectively.
Terms related to $R_{0}$ and $R_{1}$ cause more complicated
features in $I(\beta,H)$, and are neglected here for brevity.

\begin{figure} [t]
\begin{center}
\epsfig{file=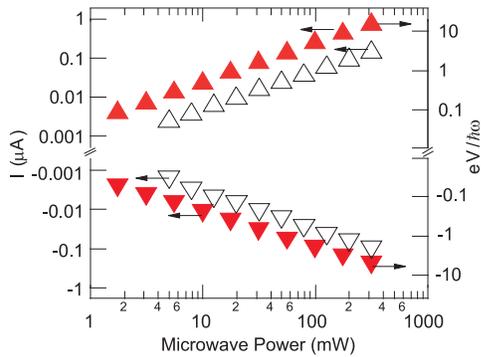,height=5 cm} \caption{(color online). The
bipolar d.c. currents generated by both the FMR (upward triangles)
and the $p$ = 2 SSW (downward triangles) depend linearly on the
microwave power over three orders of magnitude. Open marks are
maximum currents measured at 5.4 GHz on a spin dynamo with the Py
thickness $d$ = 137 nm set at $\alpha$ = -1$^{\circ}$, $\beta$ =
0. Solid marks show both the closed circuit currents (left scale)
and the open circuit voltages (right scale) measured at 6 GHz on a
second spin dynamo with $d$ = 120 nm set at $\alpha$ =
-2$^{\circ}$, $\beta$ = 0.} \label{Fig.4}
\end{center}
\end{figure}

It is clear therefore that the spin rectification effect, which
generates d.c. currents via spin dynamics, is analogous to the
optical rectification effect. The effect appears whenever spin and
charge responses mix via nonlinear responses. In semiconductors,
spin-orbit coupling may affect such a mixing \cite{Ganichev,Yang}.
In FM, the effect is nonzero due to spin-charge coupling effects.
It should be noted that a few earlier works \cite{Juretschke,
Egan} have analyzed to a great detail special features of a dc
effect induced by FMR in FM thin films. Here in this work, Eq. (2)
gives a more general solution to the generalized Ohm's equation,
and the new expression of Eq. (3) describes the spin-induced
currents in FM microstrips. Experimentally, in contrast to earlier
works \cite{Juretschke, Egan} where only a pulsed voltage signal
resonantly induced by FMR was recorded by using a high-power (up
to kW's) pulsed microwave source, we are able to directly measure
two distinct types of d.c. currents, \textit{i.e.}, $I(\alpha,H)$
induced resonantly by both FMR and SSW, and $I(\beta,H)$ caused
quasi-resonantly by the combination of AMR and the spin
excitations. These currents with different characteristics show an
intriguing bipolar symmetry which is well explained by Eq. (3).
All these striking new results are pivotal for clarifying the spin
pumping effect \cite{Costache,Azevedo}. And they are achieved due
to the exceptional technical advance of the novel spin dynamo,
which is significantly different from the devices used earlier
\cite{Juretschke, Egan}. The spin dynamo based on the lateral
seamless combination of a FM microstrip with the CPW strip lines
is also of significant technical importance, since it can be
easily integrated with modern planar technologies.

It follows from Eq. (3) that the spin rectification effect is
proportional to the microwave power. This is confirmed in Fig. 4
where the results for two spin dynamos measured over a power range
varying by three orders of magnitude are plotted. Compared to
earlier studies on semiconductors \cite{Ganichev} and FM films
\cite{Egan}, the current and voltage/power sensitivity have been
increased by three orders of magnitude, reaching values as high as
1 $\mu$A/W and 1 mV/W, respectively. Experimental data obtained
from a number of spin dynamos show that the power sensitivity
depends on the material parameters, device geometries, and
measurement configurations. The unprecedented high power
sensitivity achieved and shown in Fig. 4 is by no means the final
limit. To further increase it, one should solve the Maxwell
equations together with the Landau-Lifschitz equation under
different boundary conditions, in order to enhance the efficiency
of the microwave coupling by further optimizing the detailed
device design. It is also interesting to compare the spin dynamo
with the spin battery recently proposed for magnetic bilayers
\cite{Brataas,Tserkovnyak}. In the spin battery, a d.c. voltage
generated by FMR was predicted to occur via the spin pumping
effect, with a universal high power limit given by
$eV/\hbar\omega$. In contrast, the $eV/\hbar\omega$ ratio of the
spin dynamo reaches values approaching 18, and shows no evidence
of saturation. This indicates that the voltage contribution per
photon in the spin dynamo is more than one order of magnitude
larger than that predicted for the spin battery. We anticipate
that the difference is caused by lateral spin transport in the
spin dynamo, which was neglected for spin battery design based on
interfacial spin transport \cite{Brataas,Tserkovnyak}. Since the
characteristic length scale of the spin transport in FM is only
about a few nanometers, experimental attempts
\cite{Costache,Azevedo} to test spin battery are facing the
challenge of distinguishing effects caused by lateral and
interfacial spin transports. Results of this work provide new
solutions. It should be emphasized that the d.c signal generated
from the FM strip, as shown in Fig. 2, may reverse its sign and
change the order of its magnitude by tilting the device through
only a few tenths of a degree.

In summary, we have demonstrated a spin rectification effect which
generates d.c currents via spin wave excitations. The
unprecedented high power sensitivity and the intriguing bipolar
symmetry may enable new RF signal procession and sensor
applications utilizing spin dynamics. By highlighting the analogy
between spin and optical rectification effects, we have not only
achieved a clear understanding of the spin rectification effect in
the FM strip, but also provided a more general and consistent view
for interpreting spin-driven currents, which is currently of great
interest and have been studied in a variety of materials.

We thank G. Roy and G. Mollard for technical assistance, D.
Heitmann, U. Merkt and DFG for the loan of equipment.  N.M. is
supported by a DAAD scholarship. This work has been funded by
NSERC and URGP grants awarded to C.M.H.

\end{document}